\documentclass[12pt,tightenlines,eqsecnum,floats,aps,amsmath,amssymb,nofootinbib,prd,showpacs,preprint]{revtex4}
\usepackage{amsmath,amssymb,amsbsy,latexsym,graphicx}
\usepackage{psfrag} 
\usepackage{mathrsfs} 

\newcommand{\scri}{\mathscr{I}}

\def\R{{\mathbb R}}
\def\S{{\mathbb S}}

\def\man{{\cal M}}
\def\I{{\cal I}}

\def\Lie{{\cal L}}
\def\G{\mathbf{\Gamma}}

\def\o{\omega}
\def\Omegabf{\mathbf{\Omega}}

\def\La{\Lambda} 
\def\w{\wedge}
\def\tg{\tilde{g}}
\def\th{\tilde{h}}
\def\tE{\tilde{E}}
\def\tnabla{\tilde{\nabla}}
\def\tn{\tilde{n}}
\def\tu{\tilde{u}}
\def\te{\tilde{e}}
\def\dd{{\rm d\!I}}
\def\ww{\mathbin{\hbox{\rlap{$\mathalpha\wedge$}$\,\mathalpha\wedge$}}}
\def\bepsilon{\bar{\epsilon}}

\def\IH{{\Delta}}
\def\={\,\widehat{=}\,}
\def\l{\ell}
\def\t{\tilde}

\def\tE{\t{E}}

\def\h{\hat}
\def\u{\underline}

\def\k{\kappa}

\def\f{\frac}

\def\ba{\begin{eqnarray}}
\def\ea{\end{eqnarray}}
\def\be{\begin{equation}}
\def\ee{\end{equation}}

\preprint{\vbox{\baselineskip=12pt
 \rightline{IGPG-06/11-1}}}

\begin{document}

\title{Mechanics of higher-dimensional black holes\\ in asymptotically
anti-de Sitter space-times}

\author{Abhay\ Ashtekar${}^{1,2}$}
\email{ashtekar@gravity.psu.edu}
\author{Tomasz Pawlowski${}^1$}
\email{pawlowsk@gravity.psu.edu}
\author{Chris Van Den Broeck${}^{1,3}$}
\email{Chris.van-den-Broeck@astro.cf.ac.uk}

\affiliation{${}^1$Institute for Gravitational Physics and Geometry,\\
    Department of Physics, Penn State,
University Park, PA 16802, USA,\\
${}^2$Institute for Theoretical Physics, University of Utrecht,
    Princetonplein5, 3584 CC Utrecht, The Netherlands\\
${}^3$School of Physics and Astronomy, Cardiff University\\
Queen's Buildings, The Parade, Cardiff CF24 3AA, UK}

\begin{abstract}

We construct a covariant phase space for Einstein gravity in
dimensions $d \geq 4$ with negative cosmological constant,
describing black holes in local equilibrium. Thus, space-times
under consideration are asymptotically anti-de Sitter and admit an
inner boundary representing an isolated horizon. This allows us to
derive a first law of black hole mechanics that involves only
quantities defined quasi-locally at the horizon, \emph{without
having to assume that the bulk space-time is stationary.} The
first law proposed by Gibbons et al.~for the Kerr-AdS family
follows from a special case of this much more general first law.

\end{abstract}


\maketitle

\section{Statement of the problem}
\label{s1}

Although our universe has only four large space-time dimensions
and a positive cosmological constant, higher dimensional,
asymptotically anti-de Sitter (AdS) space-times have featured
prominently in the recent mathematical physics literature,
especially in connection with the AdS/CFT conjecture. A natural
conceptual question in this setting is whether the first law of
black hole mechanics
\be \label {1law0} \delta M = \f{\kappa}{8\pi G} \delta a + \Omega
\delta J + \Phi \delta Q\ee
continues to hold, where $M, J, Q$ and $a$ are the black hole
mass, angular momentum, charge and horizon area, and $\kappa,
\Omega, \Phi$ denote the surface gravity, angular velocity and the
electric potential. Since the last two terms have the
interpretation of work done on the black hole, similarity with the
first law of thermodynamics suggests that we interpret the first
term as the analog of $T\, \delta S$ where $T$ denotes temperature
and $S$, entropy. This in turn opens door to the fertile and
challenging field of black hole thermodynamics. Do these
considerations, which are well-established in 4 dimensions, go
through also in more general situations?

In 4 dimensions, the Kerr-Newman family of stationary black holes
admits a natural generalization to the asymptotically AdS context
and one can explicitly verify that the family continues to satisfy
(\ref{1law0}). Recall however that, to obtain the first law of
thermodynamics, one needs to assume only that the system under
consideration is in equilibrium and makes a transition to a nearby
equilibrium state; there may well be dynamical processes away from
the system. By analogy, one would physically expect the first law
of black hole mechanics to hold even in situations in which the
black hole itself is in equilibrium, although there may be time
dependence in the universe away from the black hole. This
expectation is borne out in the isolated horizon framework
\cite{abf,prl,afk,abl2} where only the \emph{intrinsic} geometry
and physical fields at the horizon are assumed to be stationary;
space-time itself need not admit any Killing field. In the
resulting extension, all quantities that enter the expression of
the first law are defined quasi-locally, at the horizon. In
particular, the mass $M$ and the angular momentum $J$ refer to the
horizon itself. In absence of global Killing vectors, they are
different from the ADM mass and angular momentum at infinity which
receive contributions also from matter and gravitational waves in
the region between the horizon and infinity. In the vacuum,
globally stationary, axi-symmetric space-times the horizon mass
and angular momentum coincides with the ADM quantities and the
first law of the isolated horizon framework reduces to that on the
AdS (or de Sitter) Kerr-Newman family.

In higher dimensions the situation is more complicated. First,
even in the stationary, axi-symmetric context, solutions
describing charged black holes are not known in dimensions higher
than 5. Hence one can not explicitly verify (\ref{1law0}) even in
this limited context. Kerr-AdS solutions, on the other hand, are
known in all dimensions \cite{hhtr,glpp}. Hence it is natural to
ask if (\ref{1law0}) holds without the last term on the right
side. Somewhat surprisingly, there has been considerable confusion
on this issue. While there is unanimity about the definition of
area, surface gravity and angular velocities, there has been lack
of agreement on the expressions of total mass and total angular
momentum. In particular, different proposals have been put forward by
Hawking, Hunter and Taylor-Robinson \cite{hhtr}, Berman and Parikh
\cite{bp}, Hawking and Reall \cite{hr}, and Awad and Johnson
\cite{aj}. However, Caldarelli, Cognola and Klemm pointed out in
$4$ dimensions and, more recently, Gibbons, Perry and Pope, in
higher dimensions that \emph{none} of these proposed conserved
quantities are compatible with the first law \cite{gpp}.

Now, in 4-dimensions and in absence of a cosmological constant, a
satisfactory covariant framework describing the asymptotic fields
and conserved quantities at spatial infinity, $i^o$, has been
available since the late seventies \cite{spi,bs}. Here, one first uses
field equations to establish certain identities involving the
asymptotic Weyl tensor and other physical fields and then uses the
asymptotic Weyl tensor to associate conserved quantities with
asymptotic symmetries. Some time ago, Ashtekar and Das (AD) showed
that this framework admits a natural generalization to higher
dimensional asymptotically AdS space-times \cite{ad}.%
\footnote{This generalization was based on the 4-dimensional
analysis of \cite{am}. In 4 dimensions, there also exist
alternative but essentially equivalent frameworks. See, e.g.,
\cite{abbot,ht}.}
Thus, there is a systematic procedure to use field equations and
asymptotic symmetries $\xi^a$ to define conserved quantities
${\cal Q}^{(\xi)}_\scri$ also in higher dimensions. More recently,
using the standard second-order Einstein-Hilbert action, Hollands,
Ishibashi and Marolf \cite{him} constructed a covariant phase
space of asymptotically AdS solutions and provided a Hamiltonian
basis for the AD quantities ${\cal Q}^{(\xi)}_\scri$. Finally,
Gibbons, Perry and Pope \cite{gpp} have shown that if one uses the
AD definitions of mass and angular momenta in Kerr-AdS solutions,
then the first law does hold for this family. This provides a
satisfactory resolution of the question of the first law for the
Kerr-AdS family in all higher dimensions.

A natural question is whether this law can be extended to
situations without global Killing fields through an isolated
horizon framework. The purpose of this paper is to answer this
question in the affirmative.

This task will require us to construct a Hamiltonian framework for
higher dimensional space-times which are asymptotically AdS
\emph{and} admit an inner boundary which is an isolated horizon
(representing a black hole in equilibrium). Thus, we will extend
three sets of constructions: i) the 4-dimensional isolated horizon
framework of \cite{afk,abl2} with a cosmological constant; ii) the
Hollands, Ishibashi, Marolf Hamiltonian framework of \cite{him}
for higher dimensional, asymptotically AdS space-times without
internal boundaries; and iii) the higher dimensional isolated
horizon framework without a cosmological constant of Korzy\'nski,
Lewandowski and Pawlowski \cite{klp}. As in \cite{afk,abl2} and
\cite{klp}, we will begin with a first order Palatini action and
construct a covariant phase space based on vielbeins and Lorentz
connections. The conserved quantities of interest will again
emerge from two surface terms, one at infinity and one at the
horizon. The terms at infinity will agree with those obtained by
Hollands, Ishibashi and Marolf \cite{him} using a covariant phase
space based on metrics, i.e., will reproduce the AD quantities
${\cal Q}^{(\xi)}_\scri$. The terms at the horizon will define the
horizon angular momenta and mass. In general, the surface
integrals at the horizon will differ from those at infinity, the
difference accounting for the energy and angular momentum in the
matter and gravitational radiation in the region between the
horizon and infinity. However, in presence of global Killing
fields, the two integrals coincide. In particular, when restricted
to the Kerr-AdS family, our first law will reduce to that obtained
by Caldarelli, Cognola and Klemm in four dimensions and by Gibbons,
Perry and Pope in higher dimensions. Because several of the necessary
techniques and intermediate steps have been discussed in 
\cite{afk,abl2,him,klp}, we will skip the corresponding details. A
more complete discussion can be found in \cite{cv}.

This paper is structured as follows. In section \ref{s2} we
construct the covariant phase space and in section \ref{s3} we
show how conserved quantities emerge as Hamiltonian functions
associated with symmetries. Detailed calculations of conserved
quantities at infinity are presented in section \ref{s4} where the
AD quantities ${\cal Q}^{(\xi)}_\scri$ are retrieved. In section
\ref{s5} we derive the expressions of energy and angular momenta
at the horizon and obtain the first law. Section \ref{s6} summarizes
the simplifications that occur in presence of global Killing vectors
and shows that the general first law obtained in section \ref{s5}
reduces to the one obtained by Gibbons, Perry and Pope \cite{gpp} in
the Kerr-AdS family.

In any Hamiltonian framework, the theory of interest has to be
specified right in the beginning. In 4 dimensions, the isolated
horizon framework has been constructed for general relativity
(with and without $\Lambda$) coupled to a large class of matter
fields, including Maxwell, dilaton, Yang-Mills, and Higgs fields.
However, as noted above, relatively little is known even about the
charged Kerr solutions in higher dimensions. Therefore, to keep
the discussion simple, in most of the paper we will restrict
ourselves to the vacuum case. Also, to avoid detours involving
technical subtleties, we will only consider non-extremal isolated
horizons. The extremal ones can be handled along the lines of
\cite{afk,abl2,abl1}. We will find it convenient to define $N := d
- 2$ where $d$ is the number of space-time dimensions. Finally, we
will set the `AdS radius' $L=\sqrt{-N(N+1)/(2\La)}$ equal to 1 in
the intermediate steps and restore it only in the final results.

\section{Covariant phase space}
\label{s2}

We wish to construct a covariant phase space $\G$ for general
relativity in $d$ space-time dimensions with a negative
cosmological constant. $\G$ consists of solutions to Einstein's
(and matter-field) equations that are asymptotically AdS and admit
a weakly isolated horizon as their inner boundary. We will work
with the first order, Palatini framework. Thus the basic variables
will be co-frames $e_a^I$ and Lorentz connections $A_a^{IJ}$ where
the lower case indices $a,b,\ldots$ denote space-time indices and
the upper case indices $I,J\ldots$, the internal (or frame)
indices. The space-time metric is defined by $g_{ab} = \eta_{IJ}
e^I_a e^J_b$ where $\eta_{IJ}$ is a fix Minkowskian metric on the
internal space $\R^d$. Throughout we will assume that $g_{ab}$ satisfies
the Einstein's equations $G_{ab} + \Lambda g_{ab} = 8\pi G
T_{ab}$, for a suitable matter stress-energy tensor $T_{ab}$
(which will be set to zero in later sections).

In the first order framework, one of the equations of motion
implies that the Lorentz connection $A_a^{IJ}$ is compatible with
the co-frame $e_a^I$ via $A^{IJ}_a = e^{bI}\nabla_a e^J_b$, where
$\nabla$ is the derivative operator compatible with $g_{ab}$.
Although the number of basic variables is larger than the metrics
$g_{ab}$ used in the second order Einstein-Hilbert framework,
detailed calculations are considerably simpler and generally more
transparent because they only involve forms and exterior calculus.

This section is divided into two parts. In the first, we specify
boundary conditions both at the outer boundary $\scri$ at infinity
and at the inner boundary representing the isolated horizon $\IH$.
In the second, we construct the covariant phase space.

\subsection{Boundary conditions}
\label{s2.1}

\begin{figure}
  \begin{center}
    \psfrag{M0}{$M$}
    \psfrag{M1}{$M_1$}
    \psfrag{M2}{$M_2$}
    \psfrag{D}{$\Delta$}
    \psfrag{tauinf}{$\scri$}
    \psfrag{S0}{$S$}
    \psfrag{Sinf}{$C$}
    \psfrag{calM}{$\mathcal{M}$}
    \includegraphics[height=3.5cm]{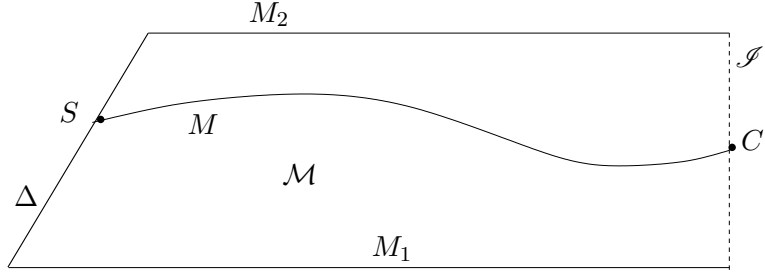}
    \caption{\small{The N+2 dimensional space-time $\man$ under
consideration has an internal boundary $\IH$ and is bounded by two
N+1-dimensional space-like surfaces $M_1$ and $M_2$. $M$ is a
partial Cauchy surface in $\man$ which intersects $\IH$ in a
compact N-dimensional surface $S$ and $\scri$ in a N dimensional
sphere $C$.}} \label{fig:phasespace}
\end{center}
\end{figure}

Let ${\man}$ be an $(N+2)$-dimensional manifold bounded by three
$(N+1)$-dimensional manifolds, $M_1$, $M_2$ and $\IH$ (see
Fig. \ref{fig:phasespace}). Dynamical fields $e_a^I$ will be
restricted such that $M_1$ and $M_2$ are space-like, while the surface
$\IH$ connecting them is null, with topology $\S\times \R$ where $\S$
is a compact $N$-dimensional manifold. Fields $(e_a^I, A_a^{IJ})$ will
be subject to boundary conditions both at infinity and at the horizon,
specified below. 

Conditions at infinity will ensure that space-times $(\man,
g_{ab})$ are asymptotically anti-de Sitter. More precisely, we
will restrict ourselves to fields $(e_a^I, A_a^{IJ})$ such that
there exists a manifold $\tilde{\man}$ with outer boundary
$\scri$, equipped with a metric $\tg_{ab}$ and a diffeomorphism
from ${\cal M}$ onto $\tilde{\cal M} - \scri$ such that:
\begin{quote}
1. There exists a smooth function $\Omega$ on $\tilde{{\cal M}}$
for which $\tg_{ab} = \Omega^2 g_{ab}$ on ${\cal M}$;\\
2. $\scri$ is topologically $\S^N \times [0,1]$, and $\Omega$
vanishes on $\scri$ but its gradient $\nabla_a \Omega$ is nowhere
vanishing on $\scri$;\\
3. On ${\cal M}$, the stress-energy tensor $T_{ab}$ is such that
$\Omega^{-N} T_{ab}$ admits a smooth limit to $\scri$. Further,
$A_a^{IJ}$ is compatible with $e^a_I$ (i.e., $A^{IJ}_a =
e^{bI}\nabla_a e^J_b$) in some neighborhood of $\scri$.
\end{quote}

In what follows, we will identify $\man$ with its image in
$\t\man$. The fall-off condition on the physical stress-energy
$T_{ab}$ is tailored to make (the energy and) the total angular
momentum of matter fields well-defined in the (stationary and)
axi-symmetric context. These boundary conditions are the standard
ones \cite{ad}. 

We will now recall from \cite{ad} and especially \cite{him} some
consequences of these boundary conditions which will play an important
role in the subsequent sections. The main idea is to use the freedom
in the conformal factor and the choice of coordinates together with
asymptotic field equations to obtain a convenient Taylor expansion
of the rescaled metric $\t{g}_{ab}$ in a neighborhood of $\scri$.
The leading order coefficients in the Taylor expansion then have
direct geometrical meaning, being related to the asymptotic Weyl
curvature. This idea was first implemented by Beig and Schmidt
\cite{bs} in the asymptotically flat, 4-dimensional context
already in the early eighties.

In the present case, it immediately follows from the field
equations (i.e., condition 3 in the definition) that $\scri$ is
time-like. Hence, there exists a neighborhood of $\scri$ in which
the $(N+1)$-dimensional surfaces $\Omega = {\rm const}$ are all
time-like. We will restrict ourselves to this neighborhood. Next,
let us set $\t{n}_a = \t\nabla_a \Omega$ so the conformally
rescaled metric has the form $\tg_{ab} = \tn_a \tn_b /(\tn
\cdot\tn)+ \th_{ab}$, where $\t{h}_{ab}$ is the induced metric (of
signature $(1, N))$ on the level surfaces of $\Omega$. One can now
Taylor-expand $\t{g}_{ab}$ in $\Omega$ and use the conformal
freedom to simplify its form. More precisely, by a judicious
modification $\Omega \mapsto e^{\alpha} \Omega$ of the conformal
factor and use of Einstein's equations, one can set $\tn^c \tn_c=1$ to
all orders in $\Omega$ so the metric takes the form
\be \tg_{ab} = \tn_a \tn_b + \th_{ab} . \ee
Next, let us Taylor expanding $\th_{ab}$ as
\be \th_{ab} = \sum_{j=0}^n \Omega^j\, \th^{(j)}_{ab}\, +\, {\cal
R}_{ab} \ee
for a sufficiently high $n$, where the remainder ${\cal R}_{ab}$ is of
the order $\mathcal{O}(\Omega^{n+1})$ (i.e., is such that even when
multiplied by $\Omega^{-(n+1)}$, it admits a smooth limit to
$\scri$). Using Einstein's equations, one arrives at a recursive
expression for the $\th^{(j)}_{ab}$. The $\th^{(j)}_{ab}$ can be
made to agree with those of the pure AdS space-time up to order
$\mathcal{O}(\Omega^{N})$. If the space-time dimension $d$ is
different from 5, the first correction, of the order
$\mathcal{O}(\Omega^{N+1})$, is given by
\be \th^{(N+1)}_{ab} = -\frac{2}{N+1} \tE_{ab}, \ee
where $\tE_{ab}$ is the appropriately rescaled, leading-order
electric part of the Weyl tensor \cite{ad}:
\be \tE_{ab} := \frac{1}{N-1}\, \Omega^{1-N}\,
\tilde{C}_{acb}{}^{d} \tilde{n}^c \tilde{n}_d. \ee
Hence the conformally rescaled metric takes the following form.
\ba \tilde{g}_{ab} &=& \t\nabla_a\Omega\,\t\nabla_b\Omega\,-
\left(1+\frac{1}{4}\Omega^2\right)^2\,\t\nabla_a t \t\nabla_b t
+ \left(1-\frac{1}{4}\Omega^2\right)^2\, S_{ab} \nonumber\\
&& -\frac{2}{N+1}\Omega^{N+1}\tilde{E}_{ab}
+ \mathcal{O}(\Omega^{N+2}), \label{unphysical2}
\ea
where $S_{ab}$ is the metric on an unit $N$ sphere. The case $d=5$
(i.e. $N=3$) is exceptional. In this case, the leading correction
to the AdS metric turns out to be:
\be \th^{(N+1)}_{ab} = -\frac{2}{N+1} \tE_{ab} - \f{1}{16}
\,\t{h}^0_{ab}, \ee
where $\t{h}^0_{ab}$ is the intrinsic metric on $\scri$ (that of
the Einstein cylinder). In this case, the asymptotic form of
$\t{g}_{ab}$ is given by
\ba \tilde{g}_{ab} &=& \t\nabla_a\Omega\,\t\nabla_b\Omega\,-
\left(1+\frac{1}{2}\Omega^2\right)\,\t\nabla_a t \t\nabla_b t
+ \left(1-\frac{1}{2}\Omega^2\right)\, S_{ab} \nonumber\\
&& -\frac{2}{N+1}\Omega^{N+1}\tilde{E}_{ab}+
\mathcal{O}(\Omega^{N+2}). \label{unphysical1} \ea
Without loss of generality one can assume that all the metrics
$\t{g}_{ab}$ of interest have the above asymptotic forms \emph{for
a fixed conformal factor $\Omega$ in a neighborhood of $\scri$} in
$\t\man$. This restriction fixes the conformal freedom in the
neighborhood and the diffeomorphism freedom to a certain
asymptotic order.%
\footnote{For details, see \cite{him}. Although this reference
assumes vacuum equations, the results quoted above hold even if
one allows matter fields whose stress energy tensor is such that
$\Omega^{-N} T_{ab}$ has a smooth limit to $\scri$.}

Next, let us consider the inner boundary $\Delta$. Now the
physical fields $(e_a^I, A_a^{IJ})$ will be assumed to be
restricted to ensure that $\Delta$ is a weakly isolated horizon.
More precisely, we assume that $\Delta$ is a null $(N+1)$-dimensional
sub-manifold of $\man$, equipped with a preferred family $[\l]$ of
future directed null normals such that:
\begin{quote}
1. $\l$ and $\l^\prime$ are in $[\l]$ if and only if there exists
a positive constant $c$ such that $\l^\prime \= c \l$ where $\=$
denotes equality at $\Delta$;\\
2. If $q_{ab}$ denotes the pull-back to $\Delta$ of the space-time
metric $g_{ab}$ (so it has signature $(0, N)$), then ${\cal L}_\l
q_{ab} \= 0$; and, if $t^a$ is any vector field tangential to
$\IH$ satisfying ${\cal L}_\ell t^a \=0$, then ${\cal L}_\l
(t^a\nabla_a \l^b) \= 0$;\\
3. \emph{On} $\IH$, the stress energy tensor satisfies $T_{ab}
\l^a \l^b \ge 0$ and $A_a^{IJ}$ is compatible with the co-frame,
(i.e., $A^{IJ}_a \= e^{bI}\nabla_a e^J_b$).
\end{quote}

These conditions capture the idea that $\Delta$ represents a
horizon in equilibrium in the following sense. First, the
condition ${\cal L}_\l q_{ab} \=0$ implies that the intrinsic
horizon metric is time-independent. In particular, it implies that
$\l^a$ is expansion free, whence all of its cross-sections have
the same area. The Raychaudhuri equation and the energy
condition then ensure $T_{ab}\l^a\l^b \=0$, i.e., that there is no
flux of matter across $\IH$. Finally, the condition ${\cal L}_\l
q_{ab} \=0$ also implies that there is a 1-form $\o_a$ on $\IH$
such that $t^a\nabla_a \l^b = t^a \o_a \l^b$ for all vectors $t^a$
tangential to $\IH$. $\o_a$ is called the \emph{rotation 1-form};
we will see in section \ref{s5} that it captures all the
information pertaining to the horizon angular momentum. The second
condition requires that the rotation 1-form also be
`time-independent'. (For further details, see
\cite{afk,abl2,klp}.)

Just as we had fixed once and for all $\Omega$ and hence $\tn^a$
in a neighborhood of $\scri$, we will now fix an equivalence class
$[\l^a]$ on $\Delta$, common to all geometries in the phase space.
As at $\I$, it is convenient to partially fix the gauge on $\IH$.
We will restrict our dynamical frame fields such that the
($N+1$)th frame field $e^a_{N+1}$ is in the equivalence class of
preferred null normals $[\l^a]$ and denote it by $\l^a$. We will
also denote the pull-back $e_{\underline{a}}^{N+1}$ of the
$(N+1)$th co-frame field $e_a^{N+1}$ by $n_a$, so that $\l^a n_a
\= -1$. Then it is easy to verify \cite{afk,abl2} that, because
$A_a^{IJ}$ is required to be compatible with the co-frame $e_a^I$,
the condition ${\cal L}_\l q_{ab} \=0$ is equivalent to
\be A_{\underline{a}}^{I\,\,N+1} \= 0 \label{Azero} \ee
for $I = 1, \ldots N$, where again the `under-bar' denotes
`pull-back to $\IH$'. Similarly, the condition $\Lie_\l \o_a \=0$
is equivalent to:
\be \Lie_{\ell} A_{\underline{a}}^{N+1\,\,N+2} \= 0. \label{LieAzero}
\ee
These facts will be used in the subsequent sections.

\subsection{The Symplectic Structure}
\label{s2.2}

The covariant phase space $\G$ consists of smooth solutions
$(e_a^I, A_a^{IJ})$ to the field equations on $\man$ which satisfy
the boundary conditions specified above.

However, to obtain the expression of the symplectic structure on
$\G$, one first suspends the field equations, considers all pairs
$(e_a^I, A_a^{IJ})$ on $\man$ satisfying the boundary conditions
and introduces the action principle. The symplectic structure
results from the second variation of the action. As in the
previous literature on isolated horizons, we will use the first
order, Palatini action. For vacuum gravity, it is given by:
\be S_P (e,A)\,  =\, -\frac{1}{16\pi G} \int_{\cal M} [F^{IJ} \w
\Sigma_{IJ} - 2\La \epsilon]\, +\, \frac{1}{16\pi G}\lim_{\Omega_o
\rightarrow 0}\int_{\Omega=\Omega_o} A^{IJ} \w \Sigma_{IJ}.
\label{Palatini} \ee
Here $\epsilon$ is the space-time volume element determined by the
frame-field,
\be \epsilon = e^1 \w e^2 \w \ldots \w e^{N+2}, \ee
$F^{IJ}$ is the curvature of $A^{IJ}_a$,
\be F^{IJ} = dA^{IJ} + A^I{}_K \w A^{KJ}, \label{curvature} \ee
and $\Sigma_{IJ}$ is defined by
\be \Sigma_{IJ} := \frac{1}{N!} \epsilon_{IJK_1 \ldots K_N}
e^{K_1} \w \ldots \w e^{K_N}, \label{Sigma} \ee
with $\epsilon_{I_1 \ldots I_{N+2}}$ the internal alternating
tensor. The integral in the last term of (\ref{Palatini}) is over
the level surfaces $\Omega=\Omega_o$ of the conformal factor
defined in the intersection of a neighborhood of $\I$ with $\man$.

The boundary term in the Palatini action, together with our
boundary conditions, ensures that the action is differentiable.
The resulting equation of motion for $A_a^{IJ}$ simply says that
the connection is determined by the co-frame $e_a^I$:
\be D\Sigma_{IJ} := d\Sigma_{IJ} - A^K{}_I \w \Sigma_{KJ} +
A^K{}_J \w \Sigma_{KI} = 0 \label{Einstein1} \ee
The equation of motion for the frame yields:
\be F^{IJ} \w \Sigma_{IJK} - 2\La\,\Sigma_K = 0, \label{Einstein2}
\ee
where the space-time $(N+2-k)$-forms $\Sigma_{I_1 \ldots I_k}$ are
straightforward generalizations of $\Sigma_{IJ}$ (defined in
(\ref{Sigma})), given by
\be \Sigma_{I_1 \ldots I_k} := \frac{1}{(N-k+2)!} \epsilon_{I_1
\ldots I_k K_{k+1} \ldots K_{N+2}} e^{K_{k+1}} \w \ldots \w
e^{K_{N+2}}. \ee
By using (\ref{Einstein1}) to express the curvature
$F_{ab}{}^{IJ}$ in terms of the frame fields $e_a^I$, it is easy
to show that (\ref{Einstein2}) reduces to the familiar field
equations:
\be G_{ab} + \Lambda g_{ab} = 0.\ee

Before embarking on the calculation of the second variation of the
action and the symplectic structure, let us note a few facts that will
be useful later. First, from the definition of  $\Sigma_{I_1, \ldots
  I_k}$ it follows that
\be \delta \Sigma_{I_1 \ldots I_k} = \delta e^L \w \Sigma_{I_1
\ldots I_k L} \label{varSigma} \ee
and
\be \chi \lrcorner \Sigma_{I_1 \ldots I_k} = \chi^L \Sigma_{I_1
\ldots I_k L} \label{wedgeSigma} \ee
where $\chi^a$ is any vector and $\chi^L = e^L_a \chi^a$. Second,
taking the variation of the second Einstein equation
(\ref{Einstein2}), we obtain a useful identity:
\be (\delta F^{IJ}) \w \Sigma_{IJK} = - F^{IJ} \w \delta e^L \w
\Sigma_{IJKL} + 2\La \, \delta e^L \w \Sigma_{KL}.
\label{linearized} \ee

We are now ready to derive the symplectic structure by considering
second variations of the action.%
\footnote{As usual, the procedure followed here actually leads to
a pre-symplectic structure: a closed 2-form on phase space which
is degenerate. The kernel corresponds to infinitesimal gauge
transformations. The physical phase space can then be obtained by
taking the quotient of the space of solutions by these gauge
transformations.}
Consider any region $\man^\prime$ in $\man$ and restrict the
Palatini action to this region. For any two vector fields
$\delta_1$ and $\delta_2$ tangent to the space of our dynamical
fields, one has
\be \label{id1} \delta_1\delta_2 S_P - \delta_2\delta_1 S_P -
[\delta_1,\delta_2]\,S_P=0 \ee
where $\delta f$ denotes the Lie derivative of the function $f$
along the vector field $\delta$ and $[\delta_1, \delta_2]$ is the
Lie bracket of the vector fields $\delta_1$ and $\delta_2$. Now,
\emph{when Einstein equations (\ref{Einstein1}) and
(\ref{Einstein2}) are satisfied}, the bulk term in the left hand
side vanishes and we have
\be \label{id2} \delta_1\delta_2 S_P - \delta_2\delta_1 S_P -
[\delta_1,\delta_2]\,S_P = \frac{1}{16\pi G}\int_{\partial
{\man^\prime}} [\delta_1 A^{IJ} \w \delta_2 \Sigma_{IJ} - \delta_2
A^{IJ} \w \delta_1 \Sigma_{IJ}]. \ee
Therefore, if we restrict ourselves to the the covariant phase
space $\G$, and restrict the vector fields $\delta_1, \delta_2$ to be
tangential to $\G$, the right side of (\ref{id2}) vanishes for any
$\man^\prime$. Hence,
\be j(\delta_1,\delta_2) = \frac{1}{16\pi G} [\delta_1 A^{IJ} \w
\delta_2 \Sigma_{IJ} - \delta_2 A^{IJ} \w \delta_1 \Sigma_{IJ}]
\ee
is a closed $(N+1)$-form on $\man$. This is the symplectic current.

Let $\man^\prime$ now be \emph{any} sub-region in $\man$, bounded
by partial Cauchy slices $M_1^\prime, M_2^\prime$ and a portion
$\IH^\prime$ of the isolated horizon $\IH$. Since $j(\delta_1,
\delta_2)$ is closed on $\man^\prime$, we have
\be \int_{M_1^\prime} j(\delta_1,\delta_2) + \int_{M_2^\prime}
j(\delta_1,\delta_2)+ \int_{\IH^\prime} j(\delta_1,\delta_2) 
+ \int_{\scri} j(\delta_1,\delta_2)  =0.
\label{jcon}\ee
The idea is to analyze these three fluxes to arrive at a conserved
2-form on the covariant phase space $\G$.

First consider the integral of $j(\delta_1,\delta_2)$ over
$\IH^\prime$. Now we can take over arguments from the isolated
horizon analysis of \cite{afk,abl2,klp} in a straightforward
manner because they are insensitive to the precise number of
space-time dimensions and the value of the cosmological constant.
We therefore simply provide the final result. Denote by
$S_1^\prime $ and $S_2^\prime$ the cross-sections of $\IH$ at
which $M_1^\prime$ and $M_2^\prime$ intersect $\IH$. Given any
point $(e_a^I, A_a^{IJ})$ of $\G$, let $\psi$ on $\IH$ be the
potential for the surface gravity $\kappa_{(\l)} := \l^a\o_a$
defined via: $\Lie_\l \psi \= \kappa_{(\l)}$ and $\psi\mid_{S_1}
=0$ where $S_1$ is the past boundary of $\IH$. Then,
\be \int_{\IH^\prime} j(\delta_1,\delta_2)\, = \, \frac{1}{8\pi
G}\, \left(\oint_{S_2^\prime} - \oint_{S_1^\prime}\right)\,\,
[\delta_1 \bepsilon\, \delta_2 \psi - \delta_2 \bepsilon\,
\delta_1 \psi], \label{jih}\ee
where $\bepsilon$ is the area element of the induced geometry on
$S_{1,2}^\prime$. All integrals in questions are well-defined
because the integrands involve smooth fields and the domains are
compact (possibly with boundary).

Next, let us consider the integrals over $M_1^\prime$ and
$M_2^\prime$. Since these manifolds are non-compact, we need to
first establish the finiteness of the flux of the symplectic
current. The idea is to express the integrands in terms of
conformally rescaled fields which are well-behaved on $\t\man$ and
carry out the integrals on $M_{1,2}^\prime \cup C_{1,2}^\prime$
where $C_{1,2}^\prime$ are the cross-sections at which
$M_{1,2}^\prime$ intersect $\scri$. Since these manifolds with
boundary are compact, the integrals are guaranteed to be
well-defined if the integrands are smooth.

Thus, for an arbitrary vector field $\delta$ in the tangent space
of $\G$, we need to express $\delta \Sigma_{IJ}$ and $\delta
A^{IJ}$ in terms of suitably conformally rescaled fields.  To
compute these, we use the expressions (\ref{unphysical1}) and
(\ref{unphysical2}) for the conformal metric, together with the
fact that since $\Omega$ is fixed on phase space, $\delta \Omega =
0$. For arbitrary $N \geq 2$ (i.e., $d \geq 4$) we then obtain
\be \delta g_{ab} = -\frac{2}{N+1} \Omega^{N+1}\delta \tE_{ab} +
\mathcal{O}(\Omega^{N+2}). \ee

We are now ready to evaluate $\delta \Sigma_{IJ}$
and $\delta A^{IJ}$. Using the definition (\ref{Sigma}),
for the former we have
\be \delta \Sigma_{IJ} = \frac{1}{(N-1)!} \epsilon_{IJK_1 \ldots
K_N} e^{K_1} \w \ldots \w e^{K_{N-1}} \w \delta e^{K_N}. \ee
Using \be \delta e^I_a = \frac{1}{2}(\delta g_{ab})\, e^{bI}
             = -\frac{1}{N+1} \Omega^N \delta\tE_{ab}\te^{bI}
                  + \mathcal{O}(\Omega^{N+1}),
\ee
together with $\te^I_a = \Omega e^I_a$ and $\delta \Omega=0$, we
arrive at
\be \delta \Sigma_{IJa_1 \ldots a_N} = -\frac{N}{N+1} \Omega
\epsilon_{IJK_1 \ldots K_N}
    \te^{K_1}_{[a_1} \ldots \te^{K_{N-1}}_{a_{N-1}}\delta \tE_{a_N]c}
\te^{c K_N} + \mathcal{O}(\Omega^2). \label{dSigma} \ee
Thus, $\delta \Sigma_{IJ}$ is of order  ${\cal O}(\Omega)$.
Computing the variation of $A^{IJ}$ is somewhat more involved; a
detailed calculation shows that
\be \delta A^{IJ}_a = \delta(e^{bI} \nabla_a e_b^J) = -2
\frac{N}{N+1} \Omega^N \delta \tE_{ab} \te^{b[I}\te^{c|J]}\tn_c
+\mathcal{O}(\Omega^{N+1}). \label{dA} \ee
Thus $\delta A^{IJ}_a$ is of order ${\cal O}(\Omega^N)$.
Collecting these two results we conclude for any two vector fields
$\delta_1$, $\delta_2$ on $\G$,
\be \delta_1 A^{IJ} \w \delta_2 \Sigma_{IJ} = \tilde {f}\,
\tilde\alpha, \label{dAdS} \ee
where $\tilde{f}$ is a function on $\man$ which asymptotically
falls off as $\mathcal{O}(\Omega^{N+1})$ and $\t\alpha$ is a
$(N+1)$-form on $\man$ which admits a smooth limit to $\scri.$ This
observation has two immediate consequences: \\
i) The integrals of the symplectic current $j$ on any partial
Cauchy surfaces $M^\prime$ is well defined;\\
ii) The flux of $j$ across $\scri$ vanishes identically.

These two facts and equations (\ref{jcon}) and (\ref{jih}) now
imply that the right side of
\be\Omegabf(\delta_1,\delta_2)\, =\, -\frac{1}{16\pi G} \int_M
[\delta_1 A^{IJ} \w \delta_2 \Sigma_{IJ} - \delta_2 A^{IJ} \w
\delta_1 \Sigma_{IJ}] \, +\, \frac{1}{8\pi G} \oint_S [\delta_1
\bepsilon \, \delta_2 \psi - \delta_2 \bepsilon \, \delta_1 \psi],
\label{symplectic} \ee
is well-defined and independent of the choice of the partial
Cauchy slice $M$ in $\man$ (where $S$ is the intersection of $M$
with $\IH$). Since the right hand side is bilinear and
anti-symmetric in the tangent vector fields $\delta_1,\, \delta_2$
on $\G$, it defines a 2-form on $\G$. The second variation
procedure ensures that it is closed (see e.g. \cite{abr}). This
is the (pre-)symplectic 2-form on our covariant phase space
$\Gamma$.

While the general procedure used above is well-known (see, e.g.
\cite{abr}), it had to be supplemented by two technical steps,
both arising from non-trivial boundary conditions. First, even
though there are no fluxes of physical quantities such as energy
or angular momentum across the inner boundary $\IH$, the flux of
the symplectic current across $\IH$ does not vanish. This fact has
to be folded-in appropriately in the analysis. As in other
isolated horizon discussions, this flux gave rise to a surface
term in the final expression of the symplectic structure. The
second technical point concerns behavior of the integrand of the
symplectic current at infinity. We showed that the fall-off is
such that the flux across $\scri$ vanishes so that there is no
surface term coming from $\scri$. We also showed that the fall-off
is such that the flux across partial Cauchy slices $M$ ---and
hence the expression of the symplectic structure itself--- is
well-defined. This step is more delicate and less transparent in
the second order framework \cite{him}.

Finally, for reasons given in Section \ref{s1}, we have restricted
ourselves to vacuum general relativity. However, inclusion of
matter fields, such as the Maxwell, Yang-Mills, and Higgs fields
will not involve new conceptual elements. To include any of them
one would add the corresponding term to the action and carry out
the second variation also of this term. In the gravitational
sector, our asymptotic analysis will go through provided the
stress energy tensor falls off so that $\Omega^{-N}T_{ab}$ admits
a smooth limit to $\scri$ ---an assumption that is necessary and
sufficient for the matter angular momentum to be finite. Thus, the
inclusion of these fields will not affect the gravitational part of
the symplectic structure. The total symplectic structure will simply
have additional terms corresponding to matter fields.

\section{Symmetries and Hamiltonians}
\label{s3}

Phase-space frameworks provide a universal procedure to arrive at
conserved quantities: they are constructed from the Hamiltonians
generating canonical transformations representing appropriate
symmetries. In particular, energy is the generator of a time
translation and angular momenta, generators of rotations. This
prescription is universal in the sense that it applies to all
physical systems ---particles, fields in Minkowski (or any stationary,
axi-symmetric) space-time, and space-time geometry itself.

Consider then smooth vector fields $\xi^a$ on $\man$ which
preserve the boundary conditions imposed in section \ref{s2.1}.
This implies in particular that $\xi^a$ are tangential to $\IH$
and admit smooth extensions to $\t\man$ which are tangential to
$\scri$. We will spell out other consequences in the next two
sections where we obtain explicit expressions for conserved
quantities at $\scri$ and a first law at $\IH$. Here, we confine
ourselves to general observations which will be needed in that
discussion.

Since we do not have any background fields, if $(e_a^I, A_a^{IJ})$
satisfies the field equations so does its image under smooth
diffeomorphisms of $\man$. Since each $\xi^a$ furthermore respects
boundary conditions, the infinitesimal diffeomorphisms it generates
define a vector field $\delta_\xi$ on $\G$, acting on $A$ and $e$
through Lie derivation along $\xi$,
\be  \delta_\xi e = \Lie_\xi e; \quad\quad \delta_\xi A = \Lie_\xi
A; \ee
$(\Lie_\xi e, \Lie_\xi A)$ automatically solve the linearized
equations. One can now ask whether this vector field $\delta_\xi$
is a phase space symmetry, i.e., if $\Lie_{\delta_\xi} \Omegabf$
vanishes identically on $\G$. The necessary and sufficient
condition for this to be the case is that there exist a function
$H^{(\xi)}$ on $\G$ such that for \emph{any} vector field $\delta$
tangent to $\G$,
\be \label{ham1} \delta H^{(\xi)} = \Omegabf(\delta,\delta_\xi).
\ee
$H^{(\xi)}$ is the Hamiltonian generating the infinitesimal
symmetry (i.e. infinitesimal canonical transformation)
$\delta_\xi$. In this section we will simplify the right side of
(\ref{ham1}) and bring it to a form that can be used directly to
obtain conserved quantities at $\scri$ and $\IH$.

Recall that the symplectic structure $\Omegabf(\delta_1,\delta_2)$
consists of a bulk term (an integral over a partial Cauchy surface
$M$) and a surface term (an integral over a cross-section of
$\IH$). (See Eq.~(\ref{symplectic}).) As shown in \cite{abl2}, if
$\xi$ is a symmetry of $\IH$, the surface term does not contribute
to $\Omegabf(\delta,\delta_\xi)$ (although it does contribute if
$\delta_\xi$ is replaced by a general vector field $\delta^\prime$
on $\G$). That argument is insensitive to the presence of a
cosmological constant or higher dimensions. Therefore, we have:
\ba \Omegabf(\delta,\delta_\xi) &=& -\frac{1}{16\pi G} \int_M
[\delta A^{IJ} \w \Lie_\xi \Sigma_{IJ} - \Lie_\xi A^{IJ} \w \delta
\Sigma_{IJ}] \nonumber\\
&=& \frac{(-1)^{N+1}}{16\pi G} \int_M [\delta \Sigma_{IJ} \w
\Lie_\xi A^{IJ} - \Lie_\xi \Sigma_{IJ} \w \delta A^{IJ}], \ea
where the rewriting in the second step was done for later
convenience. Using the Cartan identity we obtain
 \ba
\Omegabf(\delta,\delta_\xi) &=& \frac{(-1)^{N+1}}{16\pi G} \int_M
[\delta\Sigma_{IJ} \w d(\xi \lrcorner A^{IJ}) + \delta\Sigma_{IJ}
\w (\xi \lrcorner dA^{IJ})
\nonumber\\
&&\quad -d(\xi \lrcorner \Sigma_{IJ}) \w \delta A^{IJ}
       - (\xi \lrcorner d\Sigma_{IJ}) \w \delta A^{IJ}].
\ea
Next, by performing partial integration on the first and third
terms, we have:
\ba \Omegabf(\delta,\delta_\xi) &=& \frac{(-1)^{N+1}}{16\pi G}
\int_M [(-1)^{N+1} \delta d\Sigma_{IJ} \w (\xi \lrcorner A^{IJ})
 + \delta\Sigma_{IJ} \w (\xi \lrcorner dA^{IJ}) \nonumber\\
&&\quad\quad + (-1)^{N-1} (\xi \lrcorner \Sigma_{IJ}) \w \delta dA^{IJ}
- (\xi \lrcorner d\Sigma_{IJ}) \w \delta A^{IJ}] \nonumber\\
&& +\frac{(-1)^{N+1}}{16\pi G}\int_{\partial M}
[(-1)^N \delta \Sigma_{IJ} \w (\xi \lrcorner A^{IJ})
   - (\xi \lrcorner \Sigma_{IJ}) \w \delta A^{IJ}].
\ea
\emph{Note that the surface term in this (and subsequent)
expression(s) arises from the bulk term in the symplectic
structure (\ref{symplectic}) and is unrelated to the surface term
in (\ref{symplectic}).}

Next, we can use the definition of the curvature $F^{IJ}$
(\ref{curvature}) and the field equation (\ref{Einstein1}),
i.e., $D\Sigma_{IJ}=0$ to simplify the right side:
\ba \Omegabf(\delta,\delta_\xi) &=&\frac{(-1)^{N+1}}{16\pi G}
\int_M [\delta \Sigma_{IJ} \w (\xi \lrcorner F^{IJ})
  + (-1)^{N-1}(\xi \lrcorner \Sigma_{IJ}) \w \delta F^{IJ}]\nonumber\\
&&-\frac{1}{16\pi G} \int_{\partial M}
[\delta \Sigma_{IJ} \w (\xi \lrcorner A^{IJ})
  + \delta A^{IJ} \w (\xi \lrcorner \Sigma_{IJ})].
\ea

The final step is to show that the bulk term (i.e., the integral
over $M$) in this expression vanishes.  Note first that it can be
written as
\be \frac{-1}{16\pi G} \int_M[-\delta F^{IJ} \w (\xi \lrcorner
\Sigma_{IJ}) +(\xi \lrcorner F^{IJ}) \w \delta \Sigma_{IJ}].
\label{bulk} \ee
Using eq.~(\ref{wedgeSigma}) and the linearized Einstein equation
(\ref{linearized}), we have
\ba -\delta F^{IJ} \w (\xi \lrcorner \Sigma_{IJ})
&=& - \xi^K \delta F^{IJ} \w \Sigma_{IJK} \nonumber\\
&=& \xi^K F^{IJ} \w \delta e^L \w \Sigma_{IJKL}
- 2 \La \,\xi^K \delta e^L \w\ \Sigma_{KL},
\ea
where $\xi^K = e^K_a \xi^a$. Hence the integrand of (\ref{bulk})
can be written as:
\ba &&-\delta F^{IJ} \w (\xi \lrcorner \Sigma_{IJ})
+ (\xi \lrcorner F^{IJ}) \w \delta \Sigma_{IJ} \nonumber\\
&&\quad
= \delta e^L \w [- F^{IJ} \w (\xi \lrcorner \Sigma_{IJL})
+ 2 \La \,(\xi \lrcorner \Sigma_L) - (\xi \lrcorner F^{IJ}) \w \Sigma_{IJL}]
\nonumber\\
&&\quad
= -\delta e^L \w [\xi \lrcorner (\text{Einstein equation})],
\ea
where, by `Einstein equation' we mean the left hand side of
(\ref{Einstein2}). Thus, the bulk contribution to
$\Omegabf(\delta,\delta_\xi)$ indeed vanishes and the
$\Omegabf(\delta, \delta_\xi)$ reduces just to surface terms:
\ba \Omegabf(\delta,\delta_\xi) &=& -\frac{1}{16\pi G}
\int_{\partial M} [(\xi \cdot A^{IJ}) \delta \Sigma_{IJ} + \delta
A^{IJ} \w (\xi \lrcorner \Sigma_{IJ})]\nonumber \\
&=& -\frac{1}{16\pi G} \left(\oint_{C}-\oint_S\right) [(\xi \cdot
A^{IJ}) \delta \Sigma_{IJ} + \delta A^{IJ} \w (\xi \lrcorner
\Sigma_{IJ})] , \label{omegabulk1} \ea
where $C$ and $S$ are compact $N$-manifolds in which $M$ intersects
$\scri$ and $\IH$ respectively. This reduction of the expression
of $\Omegabf(\delta,\delta_\xi)$ to boundary terms is a universal
feature of field theories which are `generally covariant' i.e.
which have no background fields.%
\footnote{Thus, if $\man$ were spatially compact,
$\Omegabf(\delta,\delta_\xi)$ would vanish identically for all
$\delta$ whence $\delta_\xi$ would be in the kernel of the
symplectic structure for all smooth $\xi^a$. This is why all
diffeomorphisms represent gauge and all the corresponding
conserved quantities vanish in the spatially compact case.}

With this simplified form of $\Omegabf(\delta,\delta_\xi)$ at hand
we can return to the issue discussed in the beginning of this
section: Is $\delta_\xi$ a phase space symmetry? From (\ref{ham1})
we know that it is if and only if there exists a function
$H^{(\xi)}$ on $\G$ such that $\Omegabf(\delta,\delta_\xi) =
\delta H^{(\xi)}$ for all vector fields $\delta$ on $\G$. This
condition will be met if and only if one can `pull $\delta$ out' and
put it in front of the integrals on the right side of
(\ref{omegabulk1}). \emph{If $\xi^a$ is a vector field for which
this can be done}, then $H^{(\xi)}$ would be a difference of two
terms, one defined at $\scri$ and the other at $\IH$:
\be \label{ham2} H^{(\xi)} = {\cal Q}^{(\xi)}_\scri - {\cal
Q}^{(\xi)}_\IH. \ee
In the next two sections we will investigate conditions under
which $\delta_\xi$ is a phase space symmetry. When these
conditions are met, ${\cal Q}^{(\xi)}_\scri$ and $ {\cal
Q}^{(\xi)}_\IH$ provide us the conserved quantities at infinity
and on the horizon.

In the space-time description, the symmetry vector fields $\xi^a$
are restricted only near the boundaries where their action must
respect the boundary conditions. In the interior, they can be
\emph{arbitrary} smooth fields. In the phase space description
this is reflected by the fact that the total Hamiltonian
$H^{(\xi)}$ receives contributions only from surface terms.
Therefore, in general, symmetries and conserved quantities at
infinity are disconnected from those on the horizon (except in the
sector of the phase space admitting \emph{global} Killing fields).
Hence, on the full phase space one can calculate ${\cal
Q}^{(\xi)}_\scri$ and ${\cal Q}^{(\xi)}_\IH$ separately. To
compute the ${\cal Q}^{(\xi)}_\scri$ it will be convenient to use
vector fields $\xi^a$ that generate non-trivial symmetries at $\scri$
but vanish near $\IH$. Similarly, to calculate ${\cal Q}^{(\xi)}_\IH$
we will use vector fields that are non-trivial symmetries at the
horizon but vanish outside some neighborhood of $\IH$. The ${\cal
Q}^{(\xi)}_\scri$ will turn out to be the AD \cite{ad} quantities,
obtained in \cite{him} through a second order Hamiltonian
framework. The quantities ${\cal Q}^{(\xi)}_\IH$ will provide us
with the desired generalized first law.

In the next section we will compute the ${\cal Q}^{(\xi)}_\scri$,
while the quantities ${\cal Q}^{(\xi)}_\IH$ will be considered in
section \ref{s5}.

\section{Conserved quantities at $\scri$}
\label{s4}

Consider a vector field $\xi^a$ which has support outside of some
cylinder $\Omega = \Omega_o$ and is an asymptotic symmetry. Since
the conformal factor $\Omega$ has been fixed on our entire phase
space in a neighborhood of $\scri$, we must have ${\cal L}_\xi
\Omega =0$ in this neighborhood. Thus $\xi^a$ must be tangential
to the level surfaces of $\Omega$ there. Since $\xi^a$ must also
admit a smooth extension to $\scri$, it must be tangential to
$\scri$. Finally, since the metric at $\scri$ is also fixed to be
the AdS metric (order $\Omega^N$ in the Taylor expansion), $\xi^a$
must be the limit to $\scri$ of an AdS Killing field. Thus, the
Lie algebra of asymptotic symmetries at $\scri$ is $d(d+1)/2 =
(N+2)(N+3)/2$-dimensional. In this section we will show that for
each of these symmetry vector fields $\xi^a$ on $\man$, the vector
field $\delta_\xi$ on $\G$ is a phase space symmetry and calculate
the corresponding conserved quantity ${\cal Q}^{(\xi)}_\scri$.

Let $C$ be a cross-section of $\scri$ ---the intersection of $M$
and $\scri$. Then, from (\ref{omegabulk1}) we have
\be \Omegabf(\delta,\delta_\xi) = -\frac{1}{16\pi G} \oint_C [(\xi
\cdot A^{IJ}) \delta \Sigma_{IJ} + (-1)^{N-1}(\xi \lrcorner
\Sigma_{IJ}) \w \delta A^{IJ}]. \label{bulkC} \ee
Our goal is to show that the right side is the Lie derivative
$\delta_\xi\, {\cal Q}_{\scri}^{(\xi)}$ of a function ${\cal
Q}_{\scri}^{(\xi)}$ on $\G$.

Let us begin with the first term in the integrand,
\be (\xi \cdot A^{IJ})\,\delta \Sigma_{IJ}. \ee
From (\ref{dSigma}) we have
\be \delta \Sigma_{IJ c_1 \ldots c_N} = -\frac{N}{N+1} \Omega
\epsilon_{IJK_1 \ldots K_N}
    \te^{K_1}_{[c_1} \ldots \te^{K_{N-1}}_{c_{N-1}}\delta \tE_{c_N]d}
\te^{d K_N} + \mathcal{O}(\Omega^2). \label{indicesdSigma}
\ee
Next, let us express $A^{IJ}$ in terms of conformally rescaled
fields which are well-behaved at $\scri$:
\be A^{IJ}_a = e^{bI} \nabla_a e_b^J  = \te^{bI} \tnabla_a \te^J_b
+ \frac{2}{\Omega} \tn_b \te^{b[I} \te_a^{J]}. \ee
Collecting the two terms, we have
\be (\xi \cdot A_{IJ})\,\delta \Sigma^{IJ}{}_{c_1 \ldots c_N} =
2\frac{N}{N+1} \xi^a \tn^b \tilde{\epsilon}_{ab [c_1 \ldots
c_{N-1} |e|} \delta \tE_{c_N] d} \tg^{de} + \mathcal{O}(\Omega).
\label{xiAdSigma} \ee

Next, consider the second term in the integrand of (\ref{bulkC}):
\be (\xi \lrcorner \Sigma_{IJ}) \w \delta A^{IJ}. \ee
From (\ref{dA}) we have
\be \delta A^{IJ}_a = -2\, \frac{N}{N+1} \Omega^N \delta
\tE_{ab}\, \te^{b[I}\te^{|d|J]}\,\tn_d +\mathcal{O}(\Omega^{N+1}).
\ee
Using the definition (\ref{Sigma}) of $\Sigma_{IJ}$, we have
\be \xi^a \Sigma_{IJa c_1 \ldots c_{N-1}} := \Omega^{-N} \xi^a
\epsilon_{IJLK_1 \ldots K_{N-1}} \te^{L}_{[a} \te^{K_1}_{c_1} \w
\ldots \w e^{K_{N-1}}_{c_{N-1}]}. \ee
Combining these expressions we arrive at
\be [(\xi \lrcorner \Sigma_{IJ}) \w \delta A^{IJ}]_{c_1 \ldots
c_N} = 2(-1)^{N-1}\frac{N^2}{N+1}\, \xi^a \tn^b
\tilde{\epsilon}_{ab[c_1 \ldots c_{N-1}|e|}\, \delta \tE_{c_N]d}
\tg^{de} + \mathcal{O}(\Omega). \label{xiSigmadA} \ee

Substituting the expressions (\ref{xiAdSigma}) and (\ref{xiSigmadA})
into (\ref{bulkC}), we find
\ba \Omegabf(\delta,\delta_\xi) &=& -\frac{1}{16\pi G} \oint_C
2N\, \xi^a \tn^b\, \tilde{\epsilon}_{ab [c_1 \ldots
c_{N-1}|e|}\, \delta \tE_{c_N]d}\, \tg^{de}  \nonumber\\
&=& -\frac{1}{16\pi G} \oint_C \left[\frac{2}{(N-1)!}\, \xi^a
\tn^b\, \tilde{\epsilon}_{abc_1 \ldots c_{N-1}e} \delta
\tE_{c_Nd}\,\tg^{de} \, \tilde{\bepsilon}^{c_1 \ldots c_N}\right]
\,\tilde{\bepsilon}_{a_1 \ldots a_N} \nonumber\\
&=& -\frac{1}{8\pi G} \oint_C \delta \tE_{ab}\, \xi^a \tu^b
    \,\tilde{\bepsilon},
\ea
where $\tilde{\bepsilon}$ is again the area element on $C$ induced
by $\t{g}_{ab}$. Using $\delta \xi = 0$, and the fact that $\delta
\tu^a$ and $\delta \tilde{\bepsilon}$ vanish on $\scri$, we have
 \be \label{deltaH}
\Omegabf(\delta,\delta_\xi) = -\frac{L}{8\pi G}\,\,
\delta\left[\oint_C \tE_{ab} \xi^a \tu^b
    \,\tilde{\bepsilon}\right]
\ee
where we have reinstated the AdS radius $L$. Thus,
$\Omegabf(\delta,\delta_\xi)$
is an exact differential whence $\delta_\xi$ is a phase space
symmetry, ${\cal L}_\xi \Omegabf =0$. Since $\delta$ is an
arbitrary vector field on $\G$, (\ref{deltaH}) determines the
Hamiltonian $H^{(\xi)}$ generating this symmetry up to an additive
constant (see (\ref{ham2})). As in \cite{ad}, this constant is
determined by requiring that (\ref{ham2}) vanish in the pure AdS
space.%
\footnote{More precisely, we require that $Q^{(\xi)}_{\scri}$ should
  vanish in the limit in which the horizon area goes to zero of the
  Schwarzschild AdS family. The reason for this somewhat indirect
  condition is that the AdS space-time itself does not belong to our
  phase space since it does not admit an isolated horizon inner
  boundary.}
Then, we have:
\be H^{(\xi)} \equiv \mathcal{Q}^{(\xi)}_\scri = -\frac{L}{8\pi
G} \oint_C \tE_{ab} \xi^a \tu^b \,\tilde{\bepsilon}.
\label{ashtekardas} \ee
These are precisely the AD conserved quantities \cite{ad}.

To summarize, in this section we considered space-time vector fields
$\xi^a$ which have support only in a neighborhood of $\scri$ and
are symmetries at $\scri$, showed that $\delta_\xi$ are phase
space symmetries, and obtained the expressions of the Hamiltonians
$H^{(\xi)}$. This procedure associates with every symmetry field
$\xi^a$ on $\scri$, i.e., to every Killing field of the asymptotic
AdS metric, a conserved quantity ${\cal Q}^{(\xi)}_\scri$, thereby
providing another justification for the AD conserved quantities, now
using a first order, covariant Hamiltonian framework.

For comparison with the situation at the inner boundary $\IH$
discussed in the next section, we note that in the above
derivation we assumed that the vector field $\xi^a$ was
fixed on $\scri$ once and for all, i.e. did not vary from one
phase space point to another. Indeed, the derivation made a
crucial use of the fact that $\delta \xi$ vanishes on $\scri$. From a
space-time perspective, if $\xi^a$ is an asymptotic symmetry, so
is $k\xi^a$, where $k$ is a constant on $\scri$. A priori one can
let $k$ depend on the phase space point under consideration and
still obtain a symmetry on each individual space-time. However, in
this case $\delta_\xi$ would not be a phase space symmetry because
we would not be able to express $\Omegabf(\delta, \delta_{\xi})$
as $\delta H^{(\xi)}$. At the horizon, by contrast, we will find
that physically interesting time-translation symmetries $\xi^a$
must vary from one phase space point to another. This will make
the notion of horizon energy/mass more subtle.

\section{Conserved quantities at $\IH$ and the first law}
\label{s5}

Let us now consider symmetries and conserved quantities at the
horizon. Recall from section \ref{s2} that the definition of the
weakly isolated horizon features two fields induced directly by the
space-time metric $g_{ab}$; a preferred family $[\l^a]$ of null
normals and the intrinsic (degenerate) metric $q_{ab}$.
A vector field $\xi^a$ on $\man$ is said to be an infinitesimal
symmetry of the weakly isolated horizon $\IH$ if $\xi^a$ is
tangential to $\IH$ and satisfies $\Lie_\xi \l^a \= c\l^a,$ and
$\Lie_\xi q_{ab} \=0$ for all $\l^a \in [\l^a]$ and some positive
constant $c$. (As before $\=$ denotes `equality on $\IH$'.)

Note that there is a key difference from the situation at $\scri$.
All metrics $g_{ab}$ in our phase space $\G$ approach a
\emph{fixed} AdS metric at infinity and asymptotic symmetries are
Killing fields of this metric. The horizon $\IH$ on the other hand
is in a strong field region and the metric $q_{ab}$ varies from
one phase space point to another. Even in 4 dimensions $q_{ab}$ is
spherically symmetric in the Schwarzschild space-time and only
axi-symmetric in the Kerr space-time. Therefore even the number of
horizon symmetries is not universal on the phase space $\G$.

Clearly $\xi^a\= k\l^a$ is a horizon symmetry for any constant
$k$. However, since $q_{ab}$ can vary from one space-time to
another, in general there are no other symmetries. To introduce a
useful definition of angular momentum and obtain a first law
analogous to (\ref{1law0}), we will now restrict the phase space.%
\footnote{One can obtain first laws \cite{afk} even in absence of
these restrictions, but they are not as closely related to the
standard first law.}
First, we will assume that the topology of the inner boundary
$\IH$ is $\S^N\times \R$ and fix $N(N+1)/2$ vector fields
$\phi^a_i$ on $\IH$, satisfying the commutation relations of the
rotation group ${\rm SO(N+1)}$ and the condition $\Lie_\l \phi^a_i
\=0$. Second, will now restrict the phase space so that each
$q_{ab}$ induced on $\IH$ admits at least one $\phi_i^a$ as its
symmetry and every of its symmetries is a linear combination of
the type
\be \label{sym} \xi^a = k \l^a + \sum_i \Omega_i^{} \phi^a_i \ee
where the coefficients $k, \Omega_i$ are constants on $\IH$ (but
can vary from one phase space point to another). The constants
$\Omega_i$ are unrelated to the conformal factor used to attach
$\scri$ and, as we will see, can finally be thought of as `angular
velocities'.  In the physically interesting case of higher
dimensional Kerr solutions, $q_{ab}$ admits $[(N+1)/2]$ commuting,
rotational Killing fields where [...] stands for `integral part
of'.

With a slight abuse of notation, we will continue to denote the
restricted phase space by $\G$ and its (pulled-back) symplectic
structure by $\Omegabf$. Note that geometries in $\G$ need not
admit any Killing vectors even in a neighborhood of $\IH$; the
restriction is only on the metrics $q_{ab}$ induced \textit{on}
$\IH$.

We will now show that each $\phi^a_i$ gives rise to a conserved
angular momentum $J^i_\IH$ on the sector of the phase space which
admits a horizon symmetry $\xi^a_i$ with $\xi^a_i \= \phi^a_i$.
Let us furthermore assume for simplicity that $\xi^a$ vanishes
outside some neighborhood of $\IH$. Then Eq.~(\ref{omegabulk1})
yields:
\be \Omegabf(\delta,\delta_{\xi_i}) = \frac{1}{16\pi G} \oint_S
[(\xi_i \cdot A^{IJ}) \delta \Sigma_{IJ} + \delta A^{IJ} \w (\xi_i
\lrcorner \Sigma_{IJ})]\, , \label{omegaDelta} \ee
where $S$ is the intersection of $M$ with the horizon $\IH$. We
now use the consequences (\ref{Azero}) and (\ref{LieAzero}) of the
horizon boundary conditions to conclude
\be A^{N+1\,\,N+2}_{\underline{a}} = -\o_a \ee
where, as before, $\o_a$ the rotation potential 1-form, and
\be \Sigma_{N+1\,\,N+2\,\, \underline{a_1 \ldots a_N}} =
\bepsilon_{a_1 \ldots a_N}. \ee
Expression (\ref{omegaDelta}) then reduces to
\ba \Omegabf(\delta,\delta_{\xi_i}) &=& -\frac{1}{8\pi G} \oint_S
[(\xi_i \lrcorner \o) \,\delta \bepsilon
+ \delta \o \w (\xi_i \lrcorner \bepsilon)] \nonumber\\
&=& -\frac{1}{8\pi G} \oint_S [(\phi_i \lrcorner \o)\,\delta
\bepsilon + (\phi_i \lrcorner \delta \o)\,\bepsilon]
\nonumber\\
&=& -\frac{1}{8\pi G} \delta \oint_S [(\phi_i \lrcorner
\o)\,\bepsilon]. \ea
The right side provides us the expression of $\delta J^{i}_\IH$.
Since $\delta$ is arbitrary, this determines the angular momenta
$J^i_\IH$ up to an additive constant. We will eliminate this
freedom by a natural requirement: $J^i_\IH$ should vanish in
spherically symmetric space-times. In these space-times the pull-back
of the rotational $1$-form $\omega_a$ to any spherically symmetric
cross-section vanishes on $\IH$, whence the integral on the
right side also vanishes. Therefore, the angular momenta
$J^{i(\xi)}_\IH$ ---the Hamiltonians generating the infinitesimal
symmetry $\delta_{\xi_i}$ on $\G$--- are given by the obvious
expression
\be J_\IH^{i} := -\frac{1}{8\pi G} \oint_S (\phi_i \lrcorner
\o)\,\bepsilon. \label{angularmomenta} \ee
(\ref{angularmomenta}) brings out the reason why $\omega_a$ is
referred to as the rotation 1-form.

\emph{Remark:} For some applications, it is useful to note that
the expression (\ref{angularmomenta}) can be recast in terms of
the Weyl tensor at the horizon. Since each $\phi^a_i$ is a Killing
vector on the intrinsic metric on $S$, it is divergence-free.
Hence it admits a $(N-2)$-form potential $f_i$: $\phi_i =
{}^{\u{\star}}\, df_i$ where $\u{\star}$ denotes the Hodge-dual
on $S$.%
\footnote{We follow the standard convention: On an $n$-manifold,
the hodge-dual ${}^\star f$ of an $m$-form $f$ is an $n-m$ form
defined by: $({}^\star f)_{a_1,\ldots a_{n-m}} = \f{1}{(n-m)!}
\epsilon_{a_1,\ldots a_{n-m}}{}^{b_1\ldots b_{m}}\, f_{b_1\ldots
b_m}$.}
 Using this expression in (\ref{angularmomenta}) and
integrating by parts, one obtains:
\be J_\IH^i = \f{N+1}{8\pi G}\,\,\oint_S {}^{\u{\star}} f^{ab}_i\,
C_{abc}{}^d \l^cn_d\, \label{angularmomenta2} \bepsilon \ee
This is the generalization of the 4-dimensional expression of the
angular momentum in terms of the Weyl tensor component ${\rm
Im}\Psi_2$.

As in 4-dimensions, the definition of the horizon energy is more
subtle because we can not fix once and for all a vector field on
$\IH$ and require that the time translation symmetry coincide with
it. For, in a spherically symmetric space-time the appropriate
time-translation symmetry would be along the null generators
$\l^a$ of $\IH$, while for a rotating black hole, it would be a
linear combination of $\l^a$ and rotational symmetry-fields. Thus,
while physically interesting time translations will be horizon
symmetries, i.e., will be of the type (\ref{sym}), in general one
would expect the coefficients $k$ and $\Omega_i$ to vary from one
point on the phase space to another (although they will be constants
on the $\IH$ of any one space-time).

For later convenience, let us set $k = \kappa_{(\xi)}/\kappa_{(\l)}$,
where $\k_{(\ell)}$ is the `surface gravity' associated with the
null normal $\ell^a$ via $\ell^a\nabla_a \ell^b \= \kappa_{(\ell)}\,
\ell^b$ and $\kappa_{(\xi)}$, with the component $c\ell^a$ of $\xi^a$ along 
$\ell^a$. The symmetry field $\xi^a$ is then given by
\be \xi^a = \frac{\kappa_{(\xi)}}{\kappa_{(\ell)}}\ell^a - \sum_i
\Omega^i_{(\xi)} \phi_i^a \label{horizonsymmetry}, \ee
so that $\kappa_{(\xi)}$ is now a constant on $\IH$ in any
solution but can vary from one solution to another. Then,
Eq.~(\ref{omegabulk1}) yields:
\ba \Omegabf(\delta,\delta_\xi) &=& \frac{1}{16\pi G} \oint_S
[(\xi \cdot A^{IJ}) \delta \Sigma_{IJ} + \delta A^{IJ} \w (\xi
\lrcorner
\Sigma_{IJ})] \nonumber\\
&=& -\frac{1}{8\pi G} \oint_S [(\xi \lrcorner \o) \,\delta
\bepsilon + \delta \o \w (\xi \lrcorner \bepsilon)] \nonumber\\
&=& -\frac{1}{8\pi G} \oint_S [\kappa_{(\xi)} \delta \bepsilon -
\sum_i \Omega^i_{(\xi)} (\phi_i \lrcorner \o)\,\delta \bepsilon -
\sum_i \Omega^i_{(\xi)} (\phi_i \lrcorner \delta \o)\,\bepsilon]
\nonumber\\
&=& -\frac{\kappa_{(\xi)}}{8\pi G} \delta a_\IH
 -  \sum_i \Omega^i_{(\xi)} \delta J^{i}_\IH\,.
 \label{lookslike} \ea

Since $\kappa_{(\xi)}$ and $\Omega^i_{(\xi)}$ are allowed to be
functions on phase space, in general the right side is not an
exact differential. That is, in general $\delta_\xi$ is \textit{not} a
Hamiltonian vector field on $\G$. The necessary and sufficient
condition for this is precisely that there should exist a phase
space function, call it $E^{(\xi)}_{\IH}$, such that
\be \Omegabf(\delta,\delta_\xi) = \delta E_\IH^{(\xi)}\ee
i.e. such that
\be \delta E_\IH^{(\xi)} = \frac{\kappa_{(\xi)}}{8\pi G}\delta
a_\IH + \sum_i \Omega^i_{(\xi)} \delta J_\IH^{i}.
\label{firstlaw} \ee
Note that in any one space-time in our phase space $\G$, the
acceleration on $\IH$ of the `null part' of $\xi^a$ is given by
$\kappa_{(\xi)}$ (see (\ref{horizonsymmetry})). Therefore
$\kappa_{(\xi)}$ is the surface gravity associated with the
symmetry field $\xi^a$. Similarly, integral curves of $\xi^a$
`move' with angular velocities $\Omega^i_{(\xi)}$. These
geometrical properties of $\xi^a$ imply that (\ref{firstlaw}) has
\emph{precisely the same form as the standard first law of black
hole mechanics}, $\xi^a$ replacing the globally stationary Killing
field which need not exist.

Let us call a vector field $\xi^a$ \emph{permissible} if
$\delta_\xi$ is a Hamiltonian vector field, i.e., if
(\ref{firstlaw}) is satisfied. Now, (\ref{firstlaw}) can be
rewritten using exterior derivatives $\dd$ and wedge-products
$\ww$ on the infinite dimensional phase space $\G$:
\be 0 = \dd E_\IH^{(\xi)} = \frac{1}{8\pi G} \dd \kappa_{(\xi)}
\ww \dd a_\IH +  \sum_i \dd \Omega^i_{(\xi)} \ww \dd
J_\IH^{i}, \label{dX} \ee
This immediately implies that $\xi^a$ is permissible if and only
if its surface gravity $\kappa_{(\xi)}$ and angular velocities
$\Omega_{(\xi)}^i$, regarded as functions on the phase space $\G$,
depend only on the horizon area $a_\IH$ and angular momenta
$J^i_\IH$, and furthermore satisfy the integrability conditions:
\be \frac{\partial \kappa_{(\xi)}}{\partial
J_\IH^{i}} = 8\pi G \frac{\partial
\Omega^i_{(\xi)}}{\partial a_\IH},
\quad \text{and} \quad
\frac{\partial\Omega^i_{(\xi)}}{\partial J^k_{\IH}}
= \frac{\partial\Omega^k_{(\xi)}}{\partial J^i_{\IH}}.
\label{stringentrestriction} \ee
When these conditions are satisfied, $\delta E^{(\xi)}_\IH$ also
depends only on $a_\IH$ and $J^i_\IH$. Finally, since $\delta$ is
arbitrary, (\ref{dX}) determines $E^{(\xi)}_\IH$
up to an additive constant. This freedom is eliminated by
requiring that when $J^{i}_\IH$ all vanish, $E^{(\xi)}_\IH$
should vanish in the limit $a_\IH$ tends to zero.

Thus, the first law has been generalized from stationary
space-times to isolated horizons. In the standard first law
---e.g. the one in the Kerr-AdS family obtained in \cite{gpp}---
one uses the global time-translation Killing field to define
surface gravity and angular velocities. In the present framework,
by contrast, a typical solution in $\G$ does not admit any Killing
vector. The role of the Killing field is assumed by a horizon
symmetry (\ref{horizonsymmetry}). These symmetries exist on the
entire, \emph{infinite-dimensional} phase space, including
space-times which admit dynamical processes away from $\IH$. Thus,
(\ref{firstlaw}) represents a considerable generalization of the
standard first law. In addition, the framework brings out a deeper
significance of the first law: \emph{it is the necessary and
sufficient condition for the evolution generated by the symmetry
vector field $\xi^a$ to be Hamiltonian.} However, there is now a
first laws for each permissible vector field $\xi^a$ and using
(\ref{stringentrestriction}) it is easy to give a step by step
procedure to construct an infinite number of permissible vector
fields \cite{abl2}. In the next section we will show that, when
restricted to finite dimensional subspaces of $\G$ consisting of
solutions admitting global Killing fields, one can naturally
recover the more familiar first law from (\ref{firstlaw}).

\section{Global Symmetries}
\label{s6}

We will begin with angular momentum. Let us restrict ourselves to
solutions admitting a global rotational Killing vector
$\varphi^a$, fixed once and for all on $\man$. Denote by $\u\G$ a
(maximal) connected component of $\G$ consisting of these
axi-symmetric solutions which also contains at least one spherically
symmetric solution.%
\footnote{Because the black hole uniqueness theorem fails in
higher dimensions, $\G$ may well admit several disconnected sets
$\u\G$ (and $\t\G$) of axi-symmetric solutions (or stationary
solutions).  Sectors considered here are the most interesting ones
for us because they encompass the Kerr-AdS family.}
Then, in particular $\varphi^a$ is a symmetry both at $\scri$ and
at $\IH$. Furthermore, since $\Lie_\varphi g_{ab} =0$ on $\man$,
it follows that $(\Lie_\varphi e_a^I, \Lie_\varphi A_a^{IJ})$ is
an (internal) gauge transformation on the frames and Lorentz
connections in $\u\G$. Since these belong to the kernel of the
symplectic structure we have:
\be \u\Omegabf (\u\delta, \u{\delta}_\varphi) = 0 \ee
where $\u\Omegabf$ is the pull-back of the symplectic structure to
$\u\G$, $\u\delta$ any vector field thereon and
$\u{\delta}_\varphi$ the restriction of $\delta_\varphi$ to
$\u\G$. Therefore, from (\ref{ham2}) we must have
\be \u\delta H^{(\varphi)} \equiv \u\delta {J}^{(\varphi)}_\scri -
\u\delta {J}^{(\varphi)}_\IH =0. \ee
Thus, on $\u\G$, ${J}^{(\varphi)}_\scri$ and ${J}^{(\varphi)}_\IH$
differ just by a constant. Let us evaluate the two quantities in
any spherically symmetric space-time in $\u\G$. Since
${J}^{(\varphi)}_\scri$ can be evaluated on any cross-section $C$
of $\scri$, let us choose to evaluate it on one to which all
rotational Killing fields are tangential. Then by spherical
symmetry, the field $\t{E}_{ab}\t{u}^b$ in the integrand of the AD
conserved quantity (\ref{ashtekardas}) must be proportional to the
normal $\t{u}^a$ to $C$ within $\scri$, whence
$\t{E}_{ab}\varphi^a \t{u}^b$ must vanish. This implies that
${J}^{(\varphi)}_\scri$ must vanish on $\u\G$. Furthermore, by
construction, on a spherical symmetric space-time
${J}^{(\varphi)}_\IH$ vanishes. Therefore, the constant relating
the two conserved charges is zero and we have:
\be {J}^{(\varphi)}_\scri = {J}^{(\varphi)}_\IH. \ee

Now, the Killing vector $\varphi^a$ also gives rise to a Komar
integral:
\be {J}^{(\varphi)}_{\rm Komar} = \f{N!}{32\pi G} \oint_{\h{S}}
{}^\star d\varphi \ee
where the integral is evaluated on any N-sphere $\h{S}$ homologous
to the cross section $C$ of $\scri$ and $S$ of $\IH$ to which
$\varphi^a$ is tangential. A simple calculation shows that if
$\h{S}_1$ and $\h{S}_2$ lie on a N+1 dimensional manifold $\h{M}$
to which $\varphi^a$ is everywhere tangential, Einstein's
equations $G_{ab} + \Lambda g_{ab} =0$ imply that the Komar
integral evaluated on $\h{S}_1$ equals that evaluated on
$\h{S}_2$. In this sense ${J}^{(\varphi)}_{\rm Komar}$ is a
conserved quantity.

Denote the restriction of $\varphi^a$ to $\IH$ by $\phi$. Then, a
natural question is: What is the relation between
${J}^{(\varphi)}_{\rm Komar}$ and ${J}^{(\varphi)}_\IH$? We will
now show that they are necessarily equal. Consider a cross-section
$S$ of $\IH$ and let $n_a$ be the null normal to it satisfying
$\ell^a n_a =-1$. Extend it to $\IH$ by demanding ${\cal L}_\l n_a
\=0$. Then the definition of the rotation 1-form $\omega_a$
implies: $\omega_a = - n_b \nabla_{\u{a}} \l^b = \l^b \nabla_{b}
n_{\u{a}}$, where a bar under an index denotes the pull-back to
$\IH$. Therefore, from (\ref{angularmomenta}) we have:
\be J^{(\varphi)}_\IH \,=\, -\f{1}{8\pi G}\, \oint_S (\phi
\lrcorner \o)\,\bepsilon \,=\, \frac{1}{8\pi G} \oint_S
(\nabla_{\l} \phi) \cdot n\, \bepsilon \,=\, \f{N!}{32\pi G}
\oint_S {}^\star d\phi \ee

Next, let us explore the relation between the Komar integral
${J}^{(\varphi)}_{\rm Komar}$ and the angular momentum
${J}^{(\varphi)}_\scri$ at $\scri$. In the expression of the Komar
integral, let us express $d\varphi$ in terms of the conformal factor
$\Omega$ and rescaled fields which are smooth at $\scri$:
$\nabla_a \varphi_b = \t\nabla_{[a} (\Omega^{-2}
\hat{g}_{b]c}\t\varphi^c)$. Then, using the asymptotic forms
(\ref{unphysical2}) and (\ref{unphysical1}) of the metric
$\t{g}_{ab}$ and simplifying, one obtains:
\be J^{\varphi}_{\rm Komar} = - \f{L}{8\pi G}\oint_C
\t{E}_{ab}\t\varphi^a \t{u}^b\, \bar{\t\epsilon}\, .\ee
Thus, in presence of a global rotational Killing field, the Komar
integral $J^{(\varphi)}_{\rm Komar}$, the horizon angular momentum
$J^{(\varphi)}_\IH$  and the angular momentum
$J^{(\varphi)}_\scri$ at $\scri$ all agree, \textit{even though
each is defined using quite different fields}. In particular the
higher dimensional Kerr-AdS space-times belong to our phase space
$\G$ since they are asymptotically AdS and their event horizons
are special cases of weakly isolated horizons. In these
space-times, the horizon angular momenta $J_\IH^{i}$ agree with
the Komar integrals.

Let us next consider space-times which admit a stationary Killing
field $\xi^a$. Since $\xi^a$ is in particular a horizon symmetry,
it has the form (\ref{horizonsymmetry}) on $\IH$. Denote by $\t\G$
the connected component of $\G$ consisting of stationary solutions
which also contains the Schwarzschild-AdS family. Then arguments
completely analogous to those given above for the axi-symmetric
Killing field imply ${\cal Q}^{(\xi)}_\scri = E^{(\xi)}_\IH$.
Thus, in this case, the horizon energy coincides with the AD mass
at $\scri$ and the first law (\ref{firstlaw}) takes the standard
form:
\be \delta M = \frac{\kappa_{(\xi)}}{8\pi G}\delta a_\IH + \sum_i
\Omega^i_{(\xi)} \delta J_\IH^{i}. \label{firstlaw2} \ee
This is in particular true for the Kerr-AdS family.

We will conclude with two remarks. First, note that one can again
define a Komar integral ${\cal Q}^{(\xi)}_{\rm Komar}$ on any
$N$-sphere $S$ and, if $S_1$ and $S_2$ are boundaries of a
$(N+1)$-manifold $M$ to which $\xi^a$ is everywhere tangential, the
integrals defined on $S_1$ and $S_2$ will be equal. However, since
$\xi^a$ is stationary, manifolds $M$ to which it is tangential can
not join an $N$-sphere in the asymptotic region to that in the
interior. Therefore, the conservation law is very restricted and
not of great physical interest. This is in striking contrast with
situation vis a vis angular momentum where $M$ \emph{can} join
$N$-spheres which are in the interior to those in the asymptotic
region.%
\footnote{In both cases conservation requires that the Killing
vector be tangential to $M$ because, in the presence of a
cosmological constant, a Killing vector $K^a$ satisfies $d\,
({}^\star dK)\, \propto \,\Lambda\, K \lrcorner \epsilon$. When
$\Lambda$ vanishes, $d\, ({}^\star dK)\, =0$ and conservation
holds even when $K^a$ is not tangential to $M$.}
Finally, since the angular momentum Komar integral
${J}^{(\varphi)}_{\rm Komar}$ agrees with $J^{(\varphi)}_\scri$,
in particular it follows that ${J}^{(\varphi)}_{\rm Komar}$
vanishes in the AdS space time. The corresponding Komar integral
${\cal Q}^{(\xi)}_{\rm Komar}$ for energy diverges in the limit as
the cross-section is taken to $\scri$ even in the AdS space-time.
One could try to renormalize the Komar integral by `subtracting
out' this infinity. But straightforward subtractions lead to
ambiguous results.

Second, it is instructive to compare the above derivation of
(\ref{firstlaw2}) with that by Gibbons, Perry and Pope \cite{gpp}
for the Kerr-AdS family. In our derivation, the angular momenta
$J^i_\IH$ were defined using the rotation 1-form at the horizon
(see (\ref{angularmomenta})). The mass $M$ arose as the horizon
energy $E^{(\xi)}_\IH$. The first law (\ref{firstlaw}) only
guarantees the existence of $E^{(\xi)}_\IH$. To obtain its
explicit expression we used general facts about symplectic spaces
to argue that $E^{(\xi)}_\IH$ must equal the AD quantity ${\cal
Q}^{(\xi)}_\scri$. In \cite{gpp}, angular momenta were computed by
evaluating the Komar integrals ${J}^{(\varphi_i)}_{\rm Komar}$
using explicit Kerr-AdS metrics. As shown above, these do agree
with $J^i_\IH$ in presence of global Killing fields. The mass was
determined by integrating the first law (\ref{firstlaw2}) and
eliminating the freedom in the choice of a constant by requiring
that the mass should vanish when the parameter $m$ in the explicit
expression of the Kerr-AdS metric vanishes. They then showed that
the mass so defined agrees with the AD quantity ${\cal
Q}^{(\xi)}_\scri$ associated with the stationary Killing field
$\xi^a$. Thus, in the final picture both procedures give the same
results for angular momenta and mass and the same first law in the
Kerr-AdS family, although the starting points are very different.

\section{Discussion}
\label{s7}

In sections \ref{s2} and \ref{s3}, we constructed a covariant
phase space of asymptotically AdS solutions to Einstein equations.
While this construction is conceptually similar to that of
\cite{him}, there are two main differences: i) we allowed an
internal boundary $\IH$, a weakly isolated horizon representing a
black hole in local equilibrium; and, ii) we used a first order
framework based on vielbeins and Lorentz connections, which is
especially well suited to handling this internal boundary (and to
incorporate fermions). In section \ref{s4} we focused 
on $\scri$. Through Hamiltonian considerations, we associated a
conserved quantity ${\cal Q}^{(\xi)}_\scri$ to each symmetry
$\xi^a$ at infinity and, as in \cite{him}, showed that these agree
with the AD quantities \cite{ad} that were previously defined
using the standard asymptotic techniques \cite{spi}. In section
\ref{s5} we focused on the inner boundary $\IH$ and first obtained
expressions of angular momenta using the so-called `rotation
1-form' $\omega_a$ on $\IH$.

We then considered general horizon symmetries $\xi^a$ representing
time-translations (see (\ref{horizonsymmetry})). As in
4-dimensions \cite{afk,abl2}, now there is a qualitatively new
element: physically interesting $\xi^a$ must be allowed to vary
from one space-time to another. For example, on spherically
symmetric horizons, $\xi^a$ should point along the null normal
$\l^a$ to the horizon while in the axi-symmetric case, it should
be a linear combination of $\l^a$ and rotational symmetries
$\phi^a_i$. As a result, in general the vector field $\delta_\xi$
on the infinite dimensional phase space $\G$, defined by
infinitesimal motions along $\xi^a$ in the space-time manifold
$\man$, fails to be Hamiltonian, i.e., fails to Lie-drag the
symplectic structure. The necessary and sufficient condition for
$\delta_{\xi}$ to be Hamiltonian is precisely the generalized first
law (\ref{firstlaw}).
This is a significant generalization of the standard first law
because it does not require that space-times under consideration
be globally stationary. It also sheds new light by tying the first
law with Hamiltonian evolutions along space-time symmetry vector
fields $\xi^a$.

Finally, in section \ref{s6} we showed that in presence of
\emph{global} symmetries, the quantities defined at infinity agree
with those defined on the horizon. Furthermore, for angular
momenta, these quantities also agree with Komar integrals. This
tight relation is a non-trivial consequence of field equations
since definitions of these three different sets of quantities
involve three different sets of geometrical fields. These
relations enabled us to show that the first law for the Kerr-AdS
family obtained in \cite{gpp} results as a special case of the
much more general first law of the isolated horizon framework. In
non-stationary contexts, the framework continue to provide us with
conserved quantities associated with symmetries at $\IH$ and at
$\scri$. But now there is no simple relation between the two. In
particular, while quantities defined at $\IH$ refer only to the
horizon, i.e., to the black hole in local equilibrium, quantities
at infinity receive contributions also from dynamical fields in
the region between the horizon and infinity. It is only the
horizon quantities that feature in the generalized first law.

To summarize, we have constructed a coherent framework that
encompasses and extends three sets of results: the discussion of
the first law for the Kerr-AdS family of \cite{gpp}; the known
results on isolated horizons in four
dimensions\cite{prl,afk,abl2}, and in higher dimensions but
without a cosmological constant \cite{klp}; and the Hamiltonian
framework without internal boundaries of \cite{him}. Our detailed
considerations were restricted to the source-free case. However,
as indicated in section \ref{s2}, it should be relatively
straightforward to extend the phase space to allow matter sources
such as Maxwell, Yang-Mills and Higgs fields along the lines of
\cite{afk}.

\section*{Acknowledgments} This work was supported in part by the
NSF grant PHY-0456913, the Alexander von Humboldt Foundation, the
Kramers Chair program of the University of Utrecht and the Eberly
research funds of Penn State.


\begin{thebibliography}{99}

\bibitem{abf} Ashtekar~A, Beetle~C and Fairhurst~S 1999
  Isolated horizons: a generalization of black hole mechanics
  \textit{Class.~Quantum Grav.} {\bf 16} L1--L7\\
  Ashtekar~A, Beetle~C and Fairhurst~S 2000
  Mechanics of isolated horizons
  \textit{Class.~Quantum Grav.} {\bf 17} 253--298

\bibitem{prl} Ashtekar~A, Beetle~C, Dreyer~O, Fairhurst~S,
  Krishnan~B, Lewandowski~J and Wi\'sniewski~J 2000 Generic isolated
  horizons and their applications \textit{Phys.~Rev.~Lett.} {\bf
    85} 3564--3567

\bibitem{afk} Ashtekar~A, Fairhurst~S and Krishnan~B 2000
  Isolated horizons: Hamiltonian evolution and the first law
  \textit{Phys.~Rev.} {\bf D62} 104025

\bibitem{abl2} Ashtekar~A, Beetle~C and Lewandowski~J 2001
  Mechanics of rotating isolated horizons
  \textit{Phys.~Rev.} {\bf D64} 044016

\bibitem{hhtr} Hawking~S~W, Hunter~C~J and Taylor-Robinson~M~M 1999
  Rotation and the AdS/CFT correspondence \textit{Phys.~Rev.} {\bf
    D59} 064005

\bibitem{glpp} Gibbons~G~W, L\"{u}~H, Page~D~N and Pope~C~N 2005
  The general Kerr-de Sitter metrics in all dimensions
  \textit{J.~Geom.~Phys.} \textbf{53} 49--73

\bibitem{bp} Berman~D and Parikh~M~K 1999
  Holography and rotating AdS black holes
  \textit{Phys.~Lett.} {\bf B463} 168--173

\bibitem{hr} Hawking~S~W and Reall~H~S 2000
  Charged and rotating AdS black holes and their CFT duals
  \textit{Phys.~Rev.} {\bf D61} 024014

\bibitem{aj} Awad~A~M and Johnson~C~V 2001
  Higher dimensional Kerr-AdS black holes and the AdS/CFT
  correspondence
  \textit{Phys.~Rev.} {\bf D63} 124023

\bibitem{gpp} Caldarelli~M~M, Cognola~G, and Klemm~D 2000
  Thermodynamics of Kerr-Newman-AdS Black Holes and Conformal Field
  Theories, \textit{Class.~Quantum~Grav.} \textbf{17} 399-420 \\
  Gibbons~G~W, Perry~M~J and Pope~C~N 2005
  The first law of thermodynamics for Kerr-anti-de Sitter black
  holes \textit{Class.~Quantum~Grav.} \textbf{22} 1503-1526

\bibitem{spi} Ashtekar~A and Hansen~R~O 1978 A unified treatment of
  spatial and null infinity in general relativity: Asymptotic
  symmetries and conserved quantities at spatial infinity \textit{J.
    Math. Phys.} \textbf{19} 1542-1566\\
  Ashtekar~A 1980 Asymptotic structure of the gravitational field at
  spatiual infinity \textit{General Relativity and Gravitation:
    100 Years After the birth of Albert Einstein} ed A~Held
  (Plenum, New York)\\
  Ashtekar~A and Romano~J~D 1992 Spatial infinity as boundary of
  space-time, \textit{Class.~Quantum~Grav.} \textbf{9} 1069-1100

\bibitem{bs} Beig~R and Schmidt~B 1982 Einstein's eqautions
  near spatial infinity \textit{Commun. Math. Phys.} \textbf{87}
  65--80

\bibitem{ad} Ashtekar~A and Das~S 2000
  Asymptotically anti-de Sitter space-times: conserved quantities
  \textit{Class.~Quantum~ Grav.} {\bf 17} L17--L30

\bibitem{am} Ashtekar~A and Magnon~A 1984 Asymptotically anti-de
  Sitter space-times \textit{Class.~Quantum~Grav.} \textbf{1} L39-L44

\bibitem{abbot} Abbott~L and Deser~S 1982 Stability of gravity with
  a cosmological constant \textit{Nucl. Phys.} \textbf{B195} 76

\bibitem{ht} Henneaux~M and Teitelboim~C 1985 Asymptotically
  anti-de Sitter spaces \textit{Commun. Math. Phys.} \textbf{98}
  391--424

\bibitem{him} Hollands~S, Ishibashi~A and Marolf~D 2000
  Comparison between various notions of conserved charges in
  asymptotically AdS-space-times \textit{Class.~Quantum Grav.} {\bf
    17} L17--L30

\bibitem{klp} Korzy\'nski~M, Lewandowski~J and Pawlowski~T 2005
  Mechanics of multidimensional isolated horizons
  \textit{Class.~Quantum Grav.} {\bf 22} 2001--2016

\bibitem{cv} Van Den Broeck~C 2005 Black Holes and Neutron Stars:
  Fundamental and Phenomenological Issues \textit{Ph.D. Dissertation}
  The Pennsylvania State University

\bibitem{abl1} Ashtekar~A, Beetle~C and Lewandowski~J 2002
  Geometry of generic isolated horizons \textit{Class.~Quantum
    Grav.} {\bf 19} 1195--1225

\bibitem{lp} Lewandowski~J and Pawlowski~T 2005
  Quasi-local rotating black holes in higher dimension: geometry
  \textit{Class.~Quantum Grav.} {\bf 22} 1573--1598

\bibitem{abr} Ashtekar~A, Bombelli~L and Reula~O 1990 Covariant
  phase space of asymptotically flat gravitational fields
  \textit{Mechanics, Analysis and Geometry: 200 Years after
    Lagrange} ed M~Francaviglia and D~Holm (North Holland,
  Amsterdam)

\end{thebibliography}
\end{document}